\documentclass[12pt]{article}
\usepackage{amsmath}
\usepackage[font=small,format=plain,labelfont=bf,up,textfont=it,up]{caption}
\usepackage{graphicx}

\begin{document}

\title{\textbf{Comment on `Energy spectrum of a Dirac particle with
position-dependent mass under the influence of the Aharonov-Casher effect'\footnote{To appear in Brazilian Journal of Physics (DOI: 10.1007/s13538-020-00820-4)}}}
\date{}
\author{V. B. Mendrot\thanks{
E-mail: vitor.mendrot@unesp.br} \, and A. S. de Castro\thanks{
E-mail: antonio.castro@unesp.br} \\
\\
Departamento de F\'{\i}sica, \\
Universidade Estadual Paulista \textquotedblleft J\'{u}lio de Mesquita
Filho\textquotedblright, \\
Guaratinguet\'{a}, SP, Brazil}
\maketitle

\begin{abstract}
It is shown that the paper `Energy spectrum of a Dirac particle with
position-dependent mass under the influence of the Aharonov-Casher effect',
by Oliveira, Borges and Sousa [Braz. J. Phys. \textbf{49}, 801 (2019)], is
based on a series of ingredients clearly incorrect. \newline
\end{abstract}

\section{Introduction}

In the paper `Energy spectrum of a Dirac particle with position-dependent
mass under the influence of the Aharonov-Casher effect', recently published
in the Brazilian Journal of Physics \cite{bjp}, Oliveira, Borges and Sousa
described the planar motion of a structured neutral fermion with
position-dependent mass embedded in a ($2+1$)-dimensional world instead of a
($3+1$)-dimensional Minkowski space-time.

Possible difficulties regarding the ordering of position and momentum
operators in the Schr\"{o}dinger equation with a position-dependent mass has
nothing to do with a position-dependent mass in the Dirac equation. In the
context of the Dirac equation, position-dependent mass serves as a means for
expressing the mass of the fermion $m$ plus a position-dependent scalar
interaction, and its nonrelativistic limit is the Schr\"{o}dinger equation
for a particle of mass $m$ under the influence of an effective
position-dependent potential.

It is true that the anomalous magnetic moment of a structured neutral
fermion can be modeled by the nonminimal electromagnetic tensor interaction
in $3+1$ dimensions. Indeed, in $3+1$ dimensions the Dirac spinor has four
components and the most general interaction potential matrix without
derivative couplings can be written in terms of sixteen linearly independent
matrices arranged in terms of well defined Lorentz structures: scalar (1),
pseudoscalar (1), vector (4), pseudovector (4) and tensor (6). Nevertheless,
the little ($2+1$)-dimensional space-time is a very diverse world: the Dirac
spinor has two components and there are only four two-by-two linearly
independent matrices which can be grouped in terms well defined Lorentz structures: scalar (1) and vector (3).

To adjust knowledge gained through experience in $3+1$ dimensions to fit
knowledge in $2+1$ dimensions, and vice-versa, might be a source of danger.
For example, the four-component spinor in $3+1$ dimensions takes into
account the two energy states and the two spin states of the fermion for
given momentum.\ How to accommodate the two energy states and the two spin
states of the fermion using a two-component spinor? In 1978, de Vega pointed
out that the spin of a fermion in a $2+1$-dimensional space-time is not an
independent degree of freedom and that the state can be completely specified
by its momentum and sign of the energy \cite{veg}.

The authors of Ref. \cite{bjp}, as a starting point for their calculations
in $2+1$ dimensions, built \textquotedblleft left-handed\textquotedblright\
and \textquotedblleft right-handed\textquotedblright\ projection operators
with a \textquotedblleft chirality\textquotedblright\ operator which does
not anticommutes with all the $\gamma $-matrices appearing in the Dirac
equation. The authors of Ref. \cite{bjp}, though, making reference to a book
which treats the Dirac equation only in $3+1$ dimensions, took those
anticommutation relations for granted.

Trying to obtain a quadratic form of the Dirac equation, increasing by a
factor of two the number of solutions sought, the authors of Ref. \cite{bjp}
failed to consider the derivative property of the momentum operator.

Furthermore, without any justification, the authors of Ref. \cite{bjp}
selected one of the two possible nonequivalent representations of the $%
\gamma $-matrices.

The purpose of this comment is to present and urge reasons in opposition to
a number of lapses in judgment by the authors of Ref. \cite{bjp}. Those
slips concerning vital relations for the development of their work make
their conclusions discredited.

\section{The Dirac equation in a ($2+1$)-dimensional world}

In order to clarify our criticisms, let us begin writing the Dirac equation
in a ($2+1$)-dimensional space-time. The Dirac equation for a fermion of
mass $m$ and momentum $p^{\mu }=(p^{0},\mathbf{p})$ is written (in natural
units $\hbar =c=1$) as%
\begin{equation}
\left( \gamma ^{\mu }p_{\mu }-mI-V\right) \Psi =0,
\end{equation}%
where $\mu =0,1,2$, $p_{\mu }=i\partial _{\mu }$, $V$ is the interaction
potential matrix and $I$ is the unit matrix. The state of the system is
specified by the spinor $\Psi $. The $\gamma $-matrices satisfy the algebra $%
\{\gamma ^{\mu },\gamma ^{\nu }\}=2g^{\mu \nu }I$, where $g^{\mu \nu }$
stands for the matrix element of the Minkowski metric tensor with $%
g^{00}=-g^{ii}=1$ and $g^{\mu \nu }=0$ if $\mu \neq \nu $. The $\gamma $%
-matrices can be represented by two-by-two matrices, and the unit matrix
plus the three Pauli matrices $\sigma _{i}$ form the base of the vector
space of all $2\times 2$ matrices. In general each $\gamma ^{\mu }$
possesses four matrix elements so that we should be able to form four, and
only four, independent quantities with them. Usually, the basis elements $%
\Gamma $ are chosen in such a way that the bilinear $\Psi ^{\dag }\gamma
^{0}\Gamma \Psi $ has a definite transformation property under Lorentz
transformations. We find scalar and vector structures in $2+1$ dimensions
(in the sense of proper Lorentz transformations) in such a way that the most
general interaction potential matrix with direct (nonderivative) couplings
can be written as $V=\gamma ^{\mu }A_{\mu }+IS$. The four real quantities $%
A_{\mu }$ and $S$ are called vector and scalar potentials, respectively, due
to the transformation properties under Lorentz transformations of the
bilinear forms $\Psi ^{\dag }\gamma ^{0}\gamma ^{\mu }\Psi $ and $\Psi
^{\dag }\gamma ^{0}I\Psi $. It is instructive to note the simple way that $%
A_{\mu }$ and $S$ make their appearance from the Dirac equation for a free
fermion: $p_{\mu }\rightarrow p_{\mu }-A_{\mu }$ and $m\rightarrow m+S$. Up
to unitary transformations, two specific representations can be constructed
as $\gamma ^{0}=\sigma _{3}$ and $\boldsymbol{\gamma }=\sigma _{3}%
\boldsymbol{\sigma }$, with $\boldsymbol{\sigma }=\left( \sigma _{1},s\sigma
_{2}\right) $ and $s=\pm 1$, so that the two-component Dirac equation can be
compactly written as%
\begin{equation}
\left[ I\left( p_{0}-A_{0}\right) -{\boldsymbol{\sigma }}\cdot \left( 
\mathbf{p}-\mathbf{A}\right) -\sigma _{3}\left( m+S\right) \right] \Psi =0.
\end{equation}%
Using the identity $\sigma _{i}\sigma _{j}=\delta
_{ij}I+i\sum_{k}\varepsilon _{ijk}\sigma _{k}$, where $\delta _{ij}$ is the
Kronecker delta and $\varepsilon _{ijk}$ is the Levi-Civita symbol, we can
write $\boldsymbol{\sigma }\cdot \mathbf{A}=is\boldsymbol{\sigma }\cdot
\sigma _{3}\mathbf{a}$, where $\mathbf{a}=\left( A^{2},-A^{1}\right) $, in
such a way that we end up with \cite{asc}%
\begin{equation}
\left[ I\left( p_{0}-A_{0}\right) -\boldsymbol{\sigma }\cdot \left( \mathbf{p%
}-is\sigma _{3}\mathbf{a}\right) -\sigma _{3}\left( m+S\right) \right] \Psi
=0.
\end{equation}%
The interaction potential matrix related to the space component of a vector
potential $A^{\mu }$ has been replaced by that one with the same matrix
structure of the Dirac oscillator \cite{vil} but with a more general
potential function $\mathbf{a}$. Those two different values of $s$ give rise
to two nonequivalent representations of the $\gamma $-matrices (the
solutions of the Dirac equation are $s$-dependent as can be seen, for
example, in Ref. \cite{vil}).

\section{Criticisms}

In Ref. \cite{bjp}, with the $\gamma $-matrices expressed in terms of the
Pauli matrices, the authors referred to%
\begin{equation}
\left[ \gamma ^{\mu }p_{\mu }-m\left( \mathbf{r}\right) \right] \Psi \left(
t,\mathbf{r}\right) =0 \label{mass}
\end{equation}
as the Dirac equation of a free spin-$1/2$ particle with position-dependent
mass, but in plain terms it actually describes a fermion subject to a
position-dependent scalar interaction. See Eq. (16) in Ref. \cite{bjp} for a
particular case of $m\left( \mathbf{r}\right) $. There is no free particle described by Eq. (\ref{mass}), as might be expected.

For including the Aharonov-Casher effect, the authors of Ref. \cite{bjp}
considered a tensor potential which might be written as a space component of
a vector potential.

Serving as an essential ingredient of their work, the authors of Ref. \cite%
{bjp} defined \textquotedblleft left-handed\textquotedblright\ and
\textquotedblleft right-handed\textquotedblright\ projection operators by $%
P_{L}=\left( I-\gamma ^{5}\right) /2$ and $P_{R}=\left( I+\gamma ^{5}\right)
/2$, where, they said, $\gamma ^{5}=\sigma _{1}$. However, the cited
property $P_{R}\gamma ^{\mu }=\gamma ^{\mu }P_{L}$ would be true only if $%
\left\{ \gamma ^{5},\gamma ^{\mu }\right\} =0$. This anticommutation
relation, though, is never satisfied because there is no room for a fourth
matrix anticommutating with all the $\gamma $-matrices appearing in the
Dirac equation.

Considering the \textit{desideratum} indispensable property $P_{R}\gamma
^{\mu }=\gamma ^{\mu }P_{L}$ (it does not accord with the truth, we insist),
by defining $\Psi _{R}\left( t,\mathbf{r}\right) =P_{R}\Psi \left( t,\mathbf{%
r}\right) $ the quadratic Dirac equation should be written as%
\begin{equation}
\left[ \gamma ^{\mu }p_{\mu }-m\left( \mathbf{r}\right) \right] \frac{1}{%
m\left( \mathbf{r}\right) }\left[ \gamma ^{\mu }p_{\mu }+m\left( \mathbf{r}%
\right) \right] \Psi _{R}\left( t,\mathbf{r}\right) =0,
\end{equation}%
and not%
\begin{equation}
\left[ \gamma ^{\mu }p_{\mu }-m\left( \mathbf{r}\right) \right] \left[
\gamma ^{\mu }p_{\mu }+m\left( \mathbf{r}\right) \right] \Psi _{R}\left( t,%
\mathbf{r}\right) =0.  \label{2}
\end{equation}%
Eq. (\ref{2}) is the incorrect form found in Ref. \cite{bjp} (see Eq. (5) in Ref. \cite{bjp}). In this case, the authors of Ref. 
\cite{bjp} transgressed for not considering that the momentum operator is a
derivative operator acting on $m\left( \mathbf{r}\right) $. It seems that
they have forgotten what they said in the middle of Sec. 1 of their work:
\textquotedblleft ...the mass $m\left( \mathbf{r}\right) $ is an operator,
it does not commute more with the moment operator...\textquotedblright . As
if that were not enough, the authors supported themselves in references
([55,56]) that really do not address the problem of position-dependent mass.

The quadratic Dirac equation furnishes twice the number of the solutions
sought but the authors of Ref. \cite{bjp} did not say how to get rid of
those spurious solutions.

The authors of Ref. \cite{bjp} selected one of the two possible
nonequivalent representations of the $\gamma $-matrices but they did not
justify their choice.

\section{Final remarks}

With fulcrum on the precedent considerations, the embarrassing mistakes in
Ref. \cite{bjp} make it impossible to be confident about the conclusions
manifested there.

\section*{Acknowledgement}

Grant 2019/06734-2, S\~{a}o Paulo Research Foundation (FAPESP). Grant
09126/2019-3, Conselho Nacional de Desenvolvimento Cient\'{\i}fico e Tecnol%
\'{o}gico (CNPq), Brazil.

\end{document}